\documentclass[twocolumn,english,aps,superscriptaddress, pra,groupedaddress]{revtex4-2}

\usepackage[latin9]{inputenc}
\usepackage{amsmath}
\usepackage{amssymb}
\usepackage{graphicx}
\usepackage{babel}
\usepackage{mathrsfs}
\usepackage{amsfonts}
\usepackage{epstopdf}
\usepackage{caption}
\usepackage{multirow}

\begin{document}

\title{Thermodynamic contacts and breathing mode physics of 1D p-wave Fermi gases in the high temperature limit}

\author{Jeff Maki}
 
\affiliation{Department of Physics and HKU-UCAS Joint Institute for Theoretical and Computational Physics, \\
Guangdong-Hong Kong Joint Laboratory of Quantum Matter, \\
The University of Hong Kong, Hong Kong, China}

\date{\today}

\begin{abstract}
An important tool for understanding the effects of interactions in harmonically trapped atomic gases is the examination of their collective modes. One such mode is the breathing or monopole mode, which is special as it is constrained to occur at twice the harmonic trapping frequency when the interactions are scale invariant. When the interactions are not scale invariant, the frequency of the breathing mode will deviate from twice the trap frequency. The deviation itself depends on the thermodynamic contacts, which describe how the energy changes with the interactions. In this work I examine how the thermodynamic contacts and the breathing mode frequency of a spin-polarized one-dimensional (1D) p-wave Fermi gas depend on the 1D scattering volume, $\ell$, and the effective range, $r$, in the high temperature limit. Such dynamics can be studied in experiments and provide a tool for understanding how the dynamics depend on interactions with a finite effective range. 
\end{abstract}

\maketitle

\section{Introduction}

One fundamental type of dynamics in a many-body system is the collective mode. Most often collective modes are described using linearized hydrodynamic equations, while their finite lifetime arises from higher-order effects \cite{Nozieres, Landau}.  Although they are not exact excitations as they possess a finite lifetime, collective modes provide insights on how the dynamics of the system depend on the interactions. An excellent platform for studying this physics is an atomic gas system. The experimental control over atomic gases \cite{Feshbach_Res} allows one to study thoroughly the collective mode physics for different particle statistics, dimensions, interactions, and even in different phases \cite{Bosons_rev, Esslinger03, Fermions_rev, Kohl12, Yu12, Stringari13, Zhang17, Hu18, Stringari15, Jochim18, Vale18}.

A particular atomic gas system of recent interest is a spin-polarized Fermi gas confined to one spatial dimension (1D) \cite{Cui16, Cui17, Zhang18, Hulet20, OHara20, Iida21, Zhou21,AhmedBraum21, Sekino18, Sekino21}. Since the gas is spin-polarized, the s- or even-wave interactions are suppressed, and the leading interactions are p- or odd-wave in character. From the effective-range expansion, one expects that this gas is universal and can be described by a zero-range theory \cite{Cui16}. In other words, the physics can be described by a single low-energy scattering volume, $\ell$, while the contribution to the energetics and dynamics from the effective range, $r$, should be small and perturbative. This is in contradistinction to a 3D p-wave Fermi gas where the effective range although small, is pivotal for understanding both the energetics and dynamics \cite{Zhang15,Thywissen16, Kolck99, Kolck02, Maki20}.

For a harmonically trapped 1D spin-polarized p-wave Fermi gas, the most basic collective mode is the breathing mode. The breathing mode is unique in that it can be shown to occur at exactly twice the harmonic trap frequency and to be undamped when the interactions are scale-invariant \cite{Rosch97, Castin06}. For a zero-range 1D spin-polarized p-wave Fermi gas, scale invariance arises when the gas is at resonance, that is when the 1D scattering volume is divergent, or when the gas is non-interacting, $\ell = 0$. 

When one explicitly breaks the scale invariance either by a finite $\ell$ or by the presence of an effective range, $r$, the frequency of the breathing mode will deviate from its scale invariant value. As discussed for s-wave interactions  \cite{Hofman12, Yu12, Vale18} and for p-wave interactions in 2D \cite{Zhang17, Liu19}, the shift in the breathing mode frequency in fact depends on the thermodynamic contacts, which characterize the change in the energy of the system with respect to the low-energy scattering parameters, $\ell$ and $r$.

In this article I evaluate both the thermodynamic contacts, as well as the shift in the breathing mode frequency for a p-wave Fermi gas confined in 1D. In the high-temperature limit, one can obtain analytical results to leading order in the virial expansion, which is to leading order in $n \lambda_{th}$, where $n$ is the density and $\lambda_{th} = \sqrt{2\pi \hbar^2 /(m k_B T)}$ is the thermal de Broglie wavelength for a gas of atoms with mass $m$ and temperature $T$. The shift in the breathing mode frequency depends on both the scattering length and effective range contacts, $C_{\ell}$ and $C_{r}$, respectively. I then apply these results to experimentally applicable situations, in the hopes of future experimental studies.

The remainder of this article is organized as follows. Sec. \ref{sec:2Channel} provides the basics of the 1D p-wave Fermi gas using a two-channel model. Sec. \ref{sec:Contacts} evaluates the thermodynamic contacts using the virial expansion. Sec. \ref{sec:shift} then proceeds with a calculation of the frequency shift of the breathing mode within the virial expansion. The experimental applicability of these results is then discussed in Sec. \ref{sec:disc}. Finally, I conclude in Sec. \ref{sec:conc}.

\section{The two-channel model}
\label{sec:2Channel}

In this work I use a two-channel model for the 1D p-wave Fermi gas. The Hamiltonian is given by:

\begin{align}
H &= \sum_{k}\xi_k \psi^{\dagger}(k) \psi(k) \nonumber \\
&+ \sum_{Q} \left(\frac{1}{2}Q^2+ \nu_0 - 2\mu\right) \phi^{\dagger}(Q) \phi(Q) \nonumber \\
&+  \sum_{Q,k} \frac{gk}{2\sqrt{L}} \left[\phi(Q)\psi^{\dagger}\left(\frac{Q}{2} + k\right) \psi^{\dagger}\left(\frac{Q}{2} - k\right)+  h.c \right]
\label{eq:Hamiltonian}
\end{align}

\noindent In Eq.~(\ref{eq:Hamiltonian}), $\psi^{(\dagger)}(k)$ is the fermionic annihilation (creation) operator, $\phi^{(\dagger)}$ is the annihilation (creation) operator for the closed channel molecules, $\xi_k = k^2/2-\mu$ is the single particle dispersion, $\mu$ is the chemical potential, $\nu_0$ is the bare detuning of the molecular channel, and $g$ is the atom-molecule coupling. I also define $L$ as the length of the system, and both $\hbar$ and the bare atomic mass $m$ have been set to unity for the time being. The factor of $1/2$ in the interaction is to account for the fact spin polarized fermions are indistinguishable. 

\begin{figure}
\includegraphics[scale=0.5]{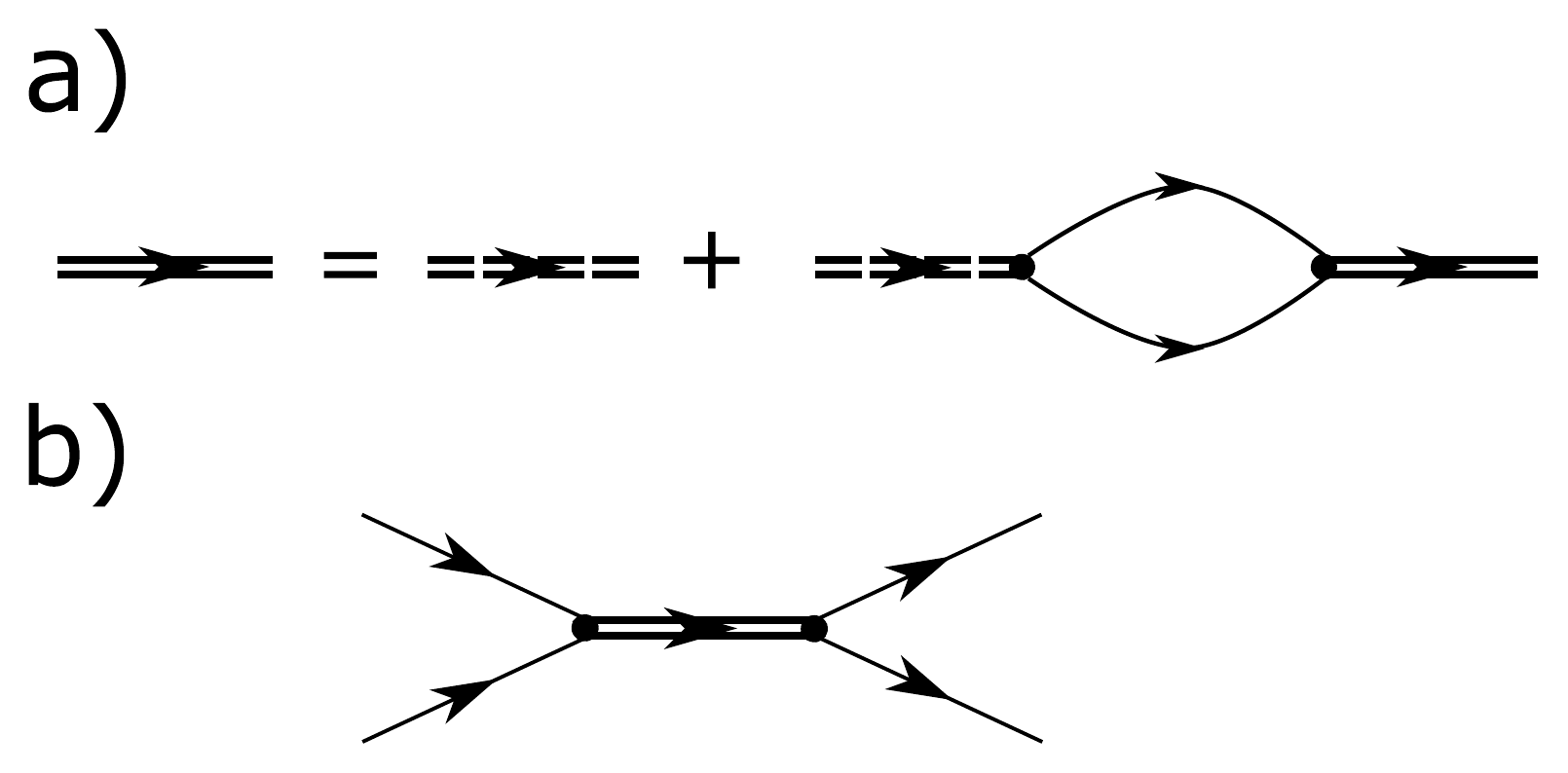}
\caption{Feynman diagrams for a) the dressed molecular propagator and for b) the T-matrix. The single lines are the free fermion propagators. The double dashed (full) lines correspond to the bare (dressed) molecular propagator. a) and b) apply both in the presence of the vacuum and the many-body background.}
\label{fig:t-matrix}
\end{figure}

Technically this theory has a ultraviolet (UV) divergence. As is custom, one can remove this divergence by examining the two-body scattering properties in the vacuum. The two-body scattering can be calculated according to Fig.~(\ref{fig:t-matrix}). Consider two spin polarized fermions with momenta, $Q/2 \pm k$, and total energy $Q_0 = Q^2/4 + k^2$.  As shown in Appendix \ref{app:T_matrix}, the $T$-matrix depends on the dressed molecular propagator, $D(Q,Q_0)$, via:

\begin{align}
\langle \frac{Q}{2} \pm k | T | \frac{Q}{2} \pm l \rangle & = k l g^2 D(Q_0-Q^2/4) \nonumber \\
\left(g^2 D(Q_0-Q^2/4)\right)^{-1} &= \left[\frac{-\nu_0}{g^2} + \frac{1}{2L}\sum_k + \frac{Q_0 - Q^2/4}{g^2} \right. \nonumber \\
& \left. - \frac{\sqrt{-Q_0 + Q^2/4-i \delta}}{4}\right]
\label{eq:vacuum_T-matrix}
\end{align}

\noindent Given the $T$-matrix one can show that the scattered relative wavefunction has the form:

\begin{align}
\psi(x) &= e^{i k x} + f_k \frac{x}{|x|}e^{i k |x|} \nonumber\\
f_k& = -i k\frac{g^2D(k^2)}{4} 
\label{eq:scattered_wf}
\end{align}

\noindent where $Q_0 - Q^2/4 = k^2$, and $f_k$ is the scattering amplitude. The scattering amplitude can then be expanded in terms of $k^2$, according to the effective range expansion:

\begin{equation}
f_k = i k \left[-\frac{1}{\ell} - r k^2 - i k +O(k^4) \right]^{-1} 
\label{eq:effective_range}
\end{equation}

\noindent This expansion is valid for low-energies, and in particular for energies $k^2 \ll r^{-2}$. 

From Eq.~(\ref{eq:vacuum_T-matrix}) one can then  identify the scattering volume, $\ell$, and the effective range, $r$:

\begin{align}
\frac{1}{4\ell} &= - \frac{\nu_0}{g^2}+ \frac{1}{2L} \sum_k  &  \frac{r}{4} &= \frac{1}{g^2}
\label{eq:scattering_parameters}
\end{align}

\noindent Here it is important to note that $r$ is strictly positive in this two-channel model, and both $\ell$ and $r$ have units of length.

\section{Thermodynamic Contacts in the High Temperature Limit}
\label{sec:Contacts}

As discussed previously \cite{Hofman12, Yu12, Vale18, Zhang17, Liu19}, the shift in the breathing mode frequency depends on the thermodynamic contacts. The thermodynamic contacts are related to the change in the energy with respect to the low-energy scattering parameters, $\ell$ and $r$:

\begin{align}
\langle C_{\ell}\rangle  &= \left\langle - \frac{\partial H}{\partial \ell^{-1}} \right\rangle \nonumber\\
&= \int_{-\infty}^{\infty} dx \frac{g^2}{4}\langle \phi^{\dagger}(x) \phi(x) \rangle \nonumber \\
\langle C_{r}\rangle  &=\left\langle - \frac{\partial H}{\partial r}\right\rangle \nonumber \\
&= \int_{-\infty}^{\infty} dx \frac{g^2}{4}\langle \phi^{\dagger}(x,t) \left(i \partial_t + \frac{\partial_x^2}{4}+ 2 \mu\right)\phi(x,t) \rangle
\label{eq:contacts}
\end{align}

\noindent The full derivation of the microscopic expressions for the contacts is presented in Appendix \ref{app:contacts}. Eq.~(\ref{eq:contacts}) is also consistent with previous definitions of the contacts \cite{Cui16}, and give consistent results for the momentum distribution and radio-frequency spectroscopy \cite{Cui16, Sekino18}.

Eq.~(\ref{eq:contacts}) can be evaluated using finite temperature field theory \cite{Mahan}. Since the contact operators are bilinear in the molecular degrees of freedom, they will be functions of the dressed molecular propagator, calculated in the presence of the many-body background:

\begin{align}
\frac{\langle C_{\ell}\rangle}{L} &= \int_{-\infty}^{\infty} \frac{dQ}{2\pi} \int_{-\infty}^{\infty} \frac{dz}{\pi} n_B(z)\Im \left[ \frac{g^2D(Q,z-i \delta)}{4}\right] \nonumber \\
\frac{\langle C_{r}\rangle}{L} &= \int_{-\infty}^{\infty} \frac{dQ}{2\pi} \int_{-\infty}^{\infty} \frac{dz}{\pi} n_B(z)\left(z - \frac{Q^2}{4}+2\mu\right)\nonumber \\
& \Im \left[ \frac{g^2D(Q,z-i \delta)}{4}\right].
\label{eq:contacts_2}
\end{align}

\noindent In Eq.~(\ref{eq:contacts_2}), $n_{B (F)}(z)$ is the Bose-Einstein (Fermi-Dirac) distribution at $\beta = 1/(k_B T)$, and the molecular propagator has been evaluated in the presence of the many-body background:

\begin{align}
&\left(\frac{g^2D(Q,z-i \delta)}{4}\right)^{-1} = \frac{1}{\ell} + \left(z-\frac{Q^2}{4}+ 2 \mu\right) r \nonumber \\
&+ \int_{-\infty}^{\infty} \frac{dk}{\pi}\ k^2  \left[\frac{1- n_F\left(\xi_{\frac{Q}{2}-k}\right) -n_F\left(\xi _{\frac{Q}{2}+k}\right)}{k^2 - \left(z-\frac{Q^2}{4}+2\mu\right) +i \delta} - \frac{1}{k^2}\right]
\label{eq:many_body_T_matrix}
\end{align}

\noindent which follows from Fig.~(\ref{fig:t-matrix}).

In the high temperature limit the chemical potential is large and negative, and thus the fugacity, $e^{\beta \mu}$, is very small: $e^{\beta \mu} \ll 1$. The expansion of Eqs.~(\ref{eq:contacts_2}-\ref{eq:many_body_T_matrix}) in terms of the fugacity is known as the virial expansion. As the contacts are related to the change in energy with respect to the scattering parameters, the contacts are proportional to $e^{2\beta\mu}$ at leading order in the virial expansion. Performing the virial expansion to this order, one obtains the following expressions for the contacts:

\begin{align}
\frac{\langle C_{\ell}\rangle}{L} &= \frac{e^{2\beta \mu}}{\sqrt{\pi \beta}} \int_{0}^{\infty} \frac{dz}{\pi} e^{-\beta z} \frac{\sqrt{z}}{\left(\frac{1}{\ell} + z r\right)^2 + z} \nonumber \\
\frac{\langle C_{r}\rangle}{L} &= \frac{e^{2\beta \mu}}{\sqrt{\pi \beta}} \int_{0}^{\infty} \frac{dz}{\pi} e^{-\beta z}z  \frac{\sqrt{z}}{\left(\frac{1}{\ell} + z r\right)^2 + z}
\label{eq:contacts_3}
\end{align}

Eq.~(\ref{eq:contacts_3}) only accounts for the scattering states contribution to the contacts and ignores the bound state contribution which only exists for $\ell >0$. As discussed in Appendix \ref{app:boundstate}, the bound-state contribution is only relevant near resonance, where the bound state energy is small but finite. Near resonance, the wavefunctions of the closed channel molecules have substantial overlap with that of the scattering states. Therefore one expects a finite number of closed channel molecules, and a finite contribution to the contacts from the bound-state. In the opposite limit of weak interactions, the two-body bound state is quite deep. If one is concerned with the upper-branch physics, the relaxation rate for pairs of individual atoms to form closed-channel molecules is quite small \cite{Feshbach_Res}. The number of closed-channel molecules is then small and the bound-state contribution to the contacts is negligible. For the simplicity of the presentation, I have relegated the molecular contribution to the contacts to Appendix \ref{app:boundstate}.
  
Eq.~(\ref{eq:contacts_3}) also suggests a relationship between the two contacts:

\begin{equation}
\left[e^{-2\beta \mu} \sqrt{\pi \beta} \frac{\langle C_r\rangle}{L}\right] = -\frac{\partial}{\partial \beta}\left[e^{-2\beta \mu} \sqrt{\pi \beta} \frac{\langle C_{\ell}\rangle}{L}\right]
\label{eq:contact_relationship}
\end{equation}

\noindent This relationship is only approximate as it neglects the many-body effects from higher orders in the virial expansion. That being said, Eq.~(\ref{eq:contact_relationship}) may provide a method for evaluating the effective range contact for a thermal gas, or to quantify the significance of many-body effects by comparing the results of Eq.~(\ref{eq:contact_relationship}) to the experimentally measured contacts.

For arbitrary $\ell$ and $r$, Eq.~(\ref{eq:contacts_3}) does not have a closed solution, and has to be determined numerically. However, there are a number of limits which allow for analytical results. First consider the gas near resonance, and with an effective range that is small compared to the thermal de Broglie wavelength: $r/\lambda_{th} \ll 1$. After restoring $\hbar$ and the atomic mass $m$, one obtains:

\begin{align}
\left.\frac{\langle C_{\ell}\rangle }{L} \right|_{\text near \ res.} &\approx 2\frac{\hbar^2n^2}{m} \left[1 - \frac{1}{\sqrt{2}} \frac{\lambda_{th}}{|\ell|} - 2 \frac{r}{\ell} + O\left(\frac{\lambda_{th}^2}{a^2} \right) \right] \nonumber\\
\left.\frac{\langle C_{r}\rangle }{L} \right|_{\text near \ res.} &\approx \frac{n^2}{\beta} \left[1 - 2 \frac{r}{\ell} + O\left(\frac{\lambda_{th}^2}{a^2} \right) \right]
\label{eq:contacts_near_res}
\end{align}

\noindent where again $\lambda_{th} = \sqrt{2\pi \hbar^2 / (m k_B T)}$ is the thermal de Broglie wavelength. I have also replaced the fugacity with the density, $n$, via: $n\lambda_{th} = e^{\beta \mu}$, which is valid to leading order in the virial expansion.

Similarly in the weakly interacting limit the contacts are given by:

\begin{align}
\left.\frac{\langle C_{\ell}\rangle }{L} \right|_{\text weak \ int.} &\approx 2\pi \frac{\hbar^2n^2}{m} \frac{\ell^2}{\lambda^2_{th}}\left[1 - 6 \pi \frac{\ell r}{\lambda_{th}^2} + O\left(\frac{\ell^2}{\lambda^2_{th}}\right)\right] \nonumber\\
\left.\frac{\langle C_{r}\rangle }{L} \right|_{\text weak \ int.} &\approx 3\pi \frac{n^2}{\beta} \frac{\ell^2}{\lambda_{th}^2}\left[1 - 10 \pi \frac{\ell r}{\lambda_{th}^2} + O\left(\frac{\ell^2}{\lambda_{th}^2} \right) \right]
\label{eq:contacts_weak_int}
\end{align}

A final important limit is to consider exactly at resonance, when $\ell^{-1} = 0$. In this case the scale invariance is solely broken by the effective range. In this limit the contacts are given by:

\begin{align}
\left.\frac{\langle C_{\ell}\rangle }{L} \right|_{\text res.} &\approx 2 \frac{\hbar^2n^2}{m} \left[1 - \pi \frac{r^2}{\lambda_{th}^2} + O\left(\frac{r^4}{\lambda_{th}^4}\right)\right] \nonumber \\
\left.\frac{\langle C_{r}\rangle }{L} \right|_{\text res.} &\approx \frac{n^2}{\beta} \left[1 - 3 \pi \frac{r^2}{\lambda_{th}^2} + O\left(\frac{r^4}{\lambda_{th}^4}\right) \right]
\label{eq:contacts_res}
\end{align}

Eqs.~(\ref{eq:contacts_near_res}-\ref{eq:contacts_res}) constitute one of the main results of this paper, and characterize the dependence of the thermodynamic contacts on the scattering parameters, $r$ and $\ell$. 

\section{Shift in the frequency of the breathing mode}
\label{sec:shift}

With the contacts in hand, it is now possible to examine the shift in the breathing mode frequency. Therefore, consider the spin-polarized 1D Fermi gas in the presence of a harmonic potential with frequency $\omega$. For this discussion I integrate out the molecular degrees of freedom, and focus only on the fermions. The total Hamiltonian is then equivalent to a one-channel model of interacting spin-polarized fermions inside a harmonic trap:

\begin{align} 
H_{trap} &= H + \omega^2 \int_{-\infty}^{\infty} dx \ \frac{x^2}{2} \psi^{\dagger}(x) \psi(x)
\label{eq:H_Trap}
\end{align}

\noindent  In writing Eq.~(\ref{eq:H_Trap}), it is important to note that the scattering properties of the one-channel Hamiltonian, $H$, are still described by the T-matrix shown in Eq.~(\ref{eq:many_body_T_matrix}).

The breathing mode for this system is an oscillatory motion associated with the moment of inertia:

\begin{equation}
\langle x^2\rangle(t) = \int dx \ x^2 n(x,t)
\label{eq:moment_of_inertia}
\end{equation}
\noindent In order to evaluate the dynamics of the moment of inertia, consider the following commutators:

\begin{align}
\left[ H, C\right] & = -i D & \left[C, D \right] &= 2 i C & \left[H,D\right] = -i \Pi
\label{eq:commutators}
\end{align}

\noindent where $C$, $D$, and $\Pi$ are the generator of conformal transformations, the generator of scale transformations, and the trace of the stress energy tensor respectively \cite{Pressure_Note}. They are defined as:

\begin{align}
C&= \int_{-\infty}^{\infty} dx \ \frac{x^2}{2} \psi^{\dagger}(x) \psi(x) \nonumber \\
D&= -i\int_{-\infty}^{\infty} dx \ \psi^{\dagger}(x) \left(x \partial_x + \frac{1}{2}\right)\psi(x) \nonumber \\
\Pi &= 2 H + \frac{1}{\ell}C_{\ell} - r C_{r}
\label{eq:operators}
\end{align}

From Eq.~(\ref{eq:commutators}) and the Heisenberg equation of motion, one can obtain the following dynamic equation for the moment of inertia:

\begin{align}
\frac{\partial^2}{\partial t^2} \langle x^2 \rangle (t) + 4 \omega^2 \langle x^2 \rangle(t) &= 4 \langle H + \omega^2 C\rangle (0)  \nonumber \\
&+  \frac{2}{\ell} \langle C_{\ell}\rangle(t) - 2 r \langle C_r\rangle(t)
\label{eq:eom_moi}
\end{align}

If one neglects the contacts, Eq.~(\ref{eq:eom_moi}) has the analytical solution:

\begin{align}
\langle x^2 \rangle(t) &= \langle x^2 \rangle(0) \left[\frac{\langle H + \omega^2 C\rangle(0)}{\omega^2 \langle x^2 \rangle(0)} \right. \nonumber \\
&\left. +\left(1 - \frac{\langle H + \omega^2 C\rangle(0)}{\omega^2 \langle x^2 \rangle(0)}\right) \cos(2\omega t)\right]
\end{align}

\noindent Thus in the absence of the contact terms, the breathing mode oscillates indefinitely at a frequency: $\omega_B = 2\omega$. This was first pointed out in Ref.~\cite{Rosch97}, and is a consequence of the non-relativistic conformal symmetry \cite{Rosch97, Castin06, Hofman12, Nishida07, Maki20b}. Physically, this situation occurs when the Hamiltonian is scale invariant, i.e. when $\ell^{-1} = 0$ and $r = 0$, or when $\ell = 0$.

The difficulty in evaluating Eq.~(\ref{eq:eom_moi}) in general is in determining the time-dependence of the thermodynamic contacts. To this end it is important to note that the breathing mode motion is dilatory in nature and can be described with a time-dependent density of the form:

\begin{equation}
n(x,t) = \frac{1}{\lambda(t)} n\left(\frac{x}{\lambda(t)}\right)
\label{eq:density_ansatz}
\end{equation}

\noindent  Such a scaling ansatz presupposes the density profile does not depend on the scattering parameters $\ell$ and $r$. In other words, Eq.~(\ref{eq:density_ansatz}) assumes that the gas follows a Boltzmann distribution for a non-interacting gas, which is valid to lowest order in the virial expansion where: $n \lambda_{th} = e^{\beta \mu}$. Since the shift in the breathing mode frequency is solely due to the contacts, which are already second order in the virial expansion, it is sufficient to assume that the dynamics of the system are also dilatory in nature.

From Eq.~(\ref{eq:density_ansatz}), the dynamics of the moment of inertia are also equivalent to a time-dependent rescaling:

\begin{equation}
\langle x^2 \rangle(t) = \lambda^2(t) \langle x^2 \rangle(0)
\label{eq:time_dep_moi}
\end{equation}

\noindent Furthermore, due to number conservation, one can show that the scaling dynamics of the density, Eq.~(\ref{eq:density_ansatz}), is equivalent to a time-dependent temperature and chemical potential: $\beta(t) = \beta \lambda^2(t)$ and $\mu(t) = \mu/\lambda^2(t)$. This allows one to explicitly evaluate the time-dependence of the contacts by simply replacing $\beta$ and $\mu$ in Eq.~(\ref{eq:contacts_3}) with their time-dependent counterparts. The result can be written in the form:

\begin{align}
\langle C_{\ell}\rangle(t) &= \frac{1}{\lambda(t)}\langle C_{\ell} \rangle \left( \frac{\lambda(t)}{\ell}, \frac{r}{\lambda(t)}\right) \nonumber \\\
\langle C_{r}\rangle(t) &= \frac{1}{\lambda^3(t)}\langle C_{r} \rangle \left( \frac{\lambda(t)}{\ell}, \frac{r}{\lambda(t)}\right)
\label{eq:time_dep_contacts}
\end{align}

This conclusion ought to be contrasted with the arguments presented in Ref.~\cite{Maki20b}. In Ref.~\cite{Maki20b}, the dilatory motion of the contacts, Eq.~(\ref{eq:time_dep_contacts}), follows from an expansion of the contacts around a scale invariant point, and is only valid to leading order in the breaking of scale invariance.  In this work the dilatory motion of the contacts follows from the trivial dynamics of the density, which is valid to leading order in the virial expansion. Thus, in this work the scaling ansatz is valid for arbitary interaction strengths provided $n\lambda_{th} \ll 1$.

Substituting Eqs.~(\ref{eq:time_dep_moi}) and (\ref{eq:time_dep_contacts}) into Eq.~(\ref{eq:eom_moi}), one obtains a second order differential equation for $\lambda(t)$:

\begin{align}
\frac{\partial^2}{\partial t^2} \lambda^2(t) &+ 4 \omega^2 \lambda^2(t) = 4\omega^2 \nonumber \\ &+\frac{2}{\ell} \frac{1}{\langle x^2 \rangle(0)}\frac{1}{\lambda(t)} \langle C_{\ell} \rangle \left(\frac{\lambda(t)}{\ell}, \frac{r}{\lambda(t)}\right) \nonumber \\
&- 2r \frac{1}{\langle x^2 \rangle(0)}\frac{1}{\lambda^3(t)} \langle C_{r} \rangle \left(\frac{\lambda(t)}{\ell}, \frac{r}{\lambda(t)}\right)  \nonumber \\
\label{eq:lambda_eom}
\end{align}

\noindent where I have used $\langle H + \omega^2 C \rangle(0) = \omega^2 \langle x^2 \rangle(0)$. The breathing mode frequency can be easily obtained by considering a small change in $\lambda^2(t)$:

\begin{equation}
\lambda^2(t) = 1 + \delta \lambda^2(t)
\label{eq:Lambda_ansatz}
\end{equation}

\noindent  Substituting Eq.~(\ref{eq:Lambda_ansatz}) into Eq.~(\ref{eq:lambda_eom}) and expanding to lowest order in $\delta \lambda^2(t)$ gives:

\begin{align}
0 &= \frac{\partial^2}{\partial t^2} \delta \lambda^2(t) + \omega_B^2 \delta \lambda^2(t)
\end{align}

\noindent which states $\delta \lambda^2(t)$ oscillates at a frequency $\omega_B$. The breathing mode frequency $\omega_B$ satisfies the following relation:

\begin{align}
\frac{\omega_B^2-4\omega^2}{\omega^2} &= 
\nonumber \\
\frac{1}{\ell} & \left[1-\frac{1}{\ell} \frac{\partial}{\partial \ell^{-1}} + r \frac{\partial}{\partial r} \right]  \frac{\langle C_{\ell}\rangle(0)}{\omega^2 \langle x^2 \rangle(0)} \nonumber \\
-r & \left[3-\frac{1}{\ell} \frac{\partial}{\partial \ell^{-1}} + r \frac{\partial}{\partial r} \right]  \frac{\langle C_{r}\rangle(0)}{\omega^2 \langle x^2 \rangle(0)}
\label{eq:breathing_mode_shift}
\end{align}

Eq.~(\ref{eq:breathing_mode_shift}) is the second main result of this work. It relates the shift of the breathing mode frequency to the thermodynamic contacts. It is straightforward to see that in the case of resonance and zero effective range, i.e. when $\ell^{-1} = 0$ and $r=0$, and in the non-interacting limit, i.e. when $\ell = 0$, $\omega_B = 2 \omega$, which is consistent with non-relativistic conformal symmetry.


In order to evaluate Eq.~(\ref{eq:breathing_mode_shift}), I employ a local density approximation (LDA) in the high-temperature limit. The details of the LDA are shown in Appendix \ref{app:LDA}. Combining the LDA with the high-temperature expansion of the contacts, Eq.~(\ref{eq:contacts_3}), one obtains the following formula for the shift of the breathing mode frequency:

\begin{align}
&\frac{\omega_B^2-4\omega^2}{\omega^2} = N  \frac{\hbar \omega}{k_B T} \nonumber \\
&\left[\frac{1}{\tilde{\ell}} \left(1- \frac{1}{\tilde{\ell}} \frac{\partial}{\partial \tilde{\ell}^{-1}} + \tilde{r} \frac{\partial}{\partial \tilde{r}}\right) \int_0^{\infty} \frac{dz}{\pi} e^{-z} \frac{\sqrt{z}}{\left(\frac{1}{\tilde{\ell}} + z \tilde{r}\right)^2 + z} \right. \nonumber \\
&\left.-\tilde{r} \left(3- \frac{1}{\tilde{\ell}} \frac{\partial}{\partial \tilde{\ell}^{-1}} + \tilde{r} \frac{\partial}{\partial \tilde{r}}\right) \int_0^{\infty} \frac{dz}{\pi} e^{-z} \frac{z^{3/2}}{\left(\frac{1}{\tilde{\ell}} + z \tilde{r}\right)^2 + z}\right]
\label{eq:breathing_mode_shift_final}
\end{align}

\noindent where $\tilde{\ell} = \ell /(\sqrt{2\pi} \lambda_{th})$, $\tilde{r} = \sqrt{2\pi} r / \lambda_{th}$, and $N \hbar \omega/ k_B T\ll 1$.

\begin{figure}
\includegraphics[scale=0.6]{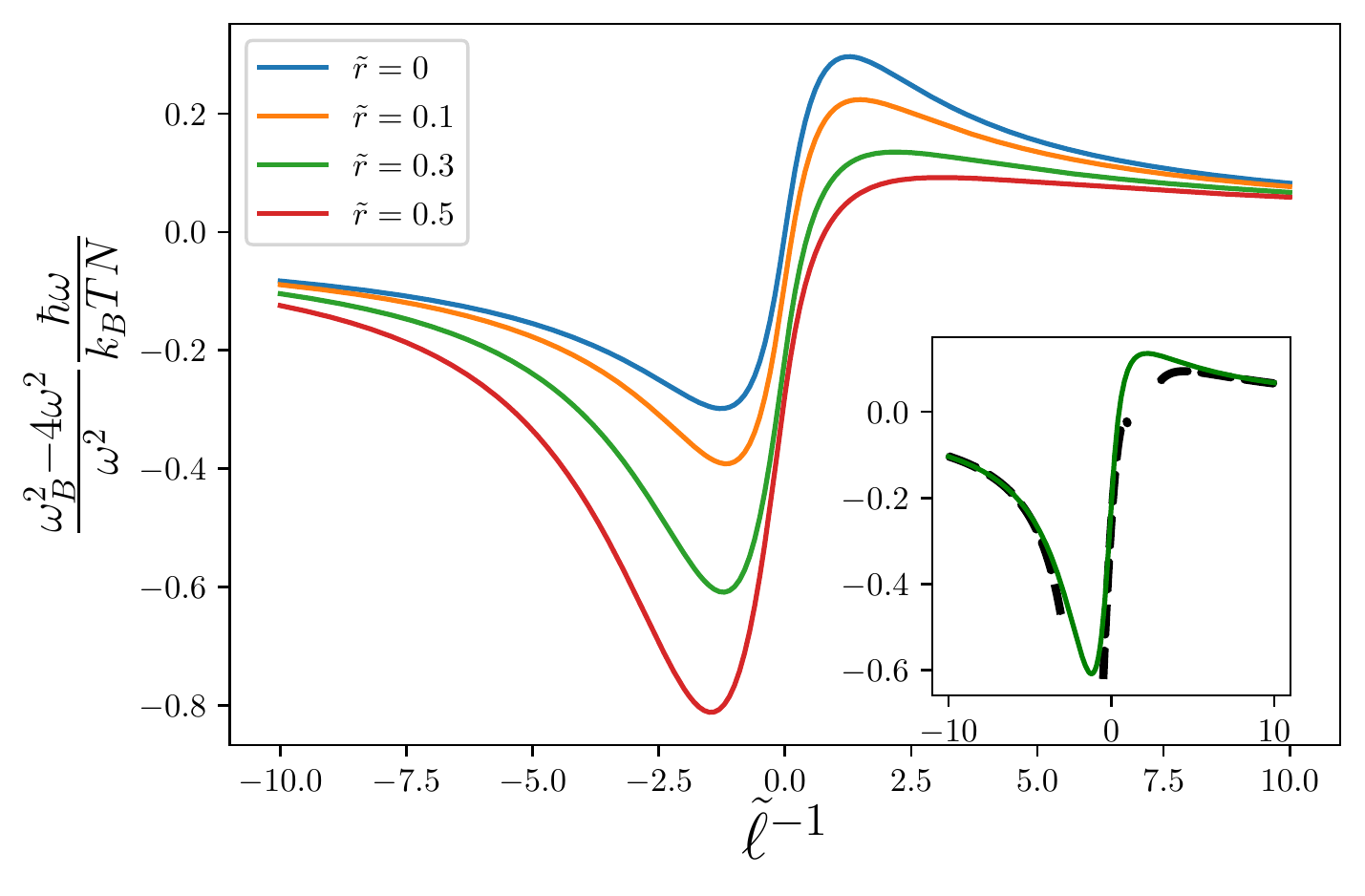}
\caption{Shift in the breathing mode frequency, Eq.~(\ref{eq:breathing_mode_shift_final}), for a spin polarized 1D p-wave Fermi gas as a function of the dimensionless scattering volume, $\tilde{\ell} = \ell/(\sqrt{2\pi}\lambda_{th})$ where $\lambda_{th}$ is the thermal de Broglie wavelength, and for various values of the dimensionless effective range, $\tilde{r} = \sqrt{2\pi} r /\lambda_{th}$. This figure disregards the molecular contribution. The inset shows the shift in the breathing mode frequency for $\tilde{r} = 0.3$ (green solid line), and the asymptotic expansions from Eqs.~(\ref{eq:shift_res}-\ref{eq:shift_weak}) (black dashed lines).}
\label{fig:breathing_mode_ell}
\end{figure}


Eq.~(\ref{eq:breathing_mode_shift_final}), has to be evaluated numerically, but one can obtain analytical expressions for the shift of the breathing mode frequency near the resonantly interacting limit:

\begin{equation}
\left. \frac{\omega_B^2-4\omega^2}{\omega^2}\right|_{\text near \ res.} \approx \frac{\hbar \omega N}{k_B T} \left[\frac{1}{\tilde{\ell} \sqrt{\pi}} \left(1-\frac{2 \tilde{r}}{\tilde{\ell}}\right) - \frac{3 \tilde{r}}{2\sqrt{\pi}}\right]
\label{eq:shift_res}
\end{equation}

\noindent and for weak interactions:

\begin{equation}
\left. \frac{\omega_B^2-4\omega^2}{\omega^2}\right|_{\text weak \ int.} \approx \frac{\hbar \omega N}{k_B T}\frac{3 \tilde{\ell}}{2\sqrt{\pi}}\left(1- \frac{15}{2}\tilde{r}\tilde{\ell}\right)
\label{eq:shift_weak}
\end{equation}

For a more complete picture, consider Fig.~(\ref{fig:breathing_mode_ell}). In this figure the shift in the breathing mode frequency is presented as a function of $\tilde{\ell}^{-1}$ for various values of the effective range, $\tilde{r}$. One prominent feature of Fig.~(\ref{fig:breathing_mode_ell}) is that the shift in the breathing mode becomes more negative as the effective range becomes larger. Conversely, the contribution due to the scattering volume, $\tilde{\ell}$, depends on the sign of $\tilde{\ell}$. This can be qualitatively understood by examining Eq.~(\ref{eq:eom_moi}) and the pressure, $P$. As shown in Appendix \ref{app:contacts}, the pressure is given by:

\begin{equation}
P L = \langle \Pi \rangle = 2 \langle H \rangle + \frac{1}{\ell} \langle C_{\ell} \rangle - r \langle C_r\rangle
\end{equation}

\noindent Since the contacts in Eq.~(\ref{eq:contacts_3}) are positive semi-definite, the shift in the pressure from its scale invariant value due to the scattering volume is positive (negative) for positive (negative) $\ell$, which is a signature of repulsive (attractive) interactions. On the other hand, the contribution due to a finite effective range, $r$, is always negative and hence attractive in nature.  This shift in the pressure is directly related to the shift in the breathing mode frequency via Eq.~(\ref{eq:breathing_mode_shift}), and explains the general features of Fig.~(\ref{fig:breathing_mode_ell}).

One can also see that the shift in the breathing mode frequency is finite at resonance due to the presence of the effective range. Away from resonance, the the shift in the breathing mode frequency is more substantial for $\tilde{\ell}<0$, as some cancellation occurs for $\tilde{\ell} >0$. In the weakly interacting limit, the contribution from the effective range becomes minuscule in comparison to the leading contribution governed by $\tilde{\ell}$,  which reinforces the idea that the effective range merely gives a perturbative correction. 


\section{Discussion}
\label{sec:disc}

The results obtained thus far are for strictly one-dimensional systems. Realistically, however, the one-dimensional physics described above is a low-energy approximation for a three-dimensional spin-polarized Fermi gas placed in a cylindrical trapping potential with tight radial confinement. The low-energy scattering volume, $\ell$, and the effective range, $r$, in the ground state of the radial confinement are related to their 3D counterparts, $v$ and $R$ respectively, via \cite{Cui17,Jiang15, Zhou21}:

\begin{align}
\frac{1}{\ell} &= \frac{a_{\perp}^2}{6}\left( \frac{1}{v} + \frac{2}{R a_{\perp}^2}\right) - \frac{2}{a_{\perp}} \zeta(-1/2)  \nonumber \\
r &= \frac{a_{\perp}^2}{6R} + \frac{a_{\perp}}{4} \zeta(1/2)
\label{eq:cir}
\end{align}

\noindent where $a_{\perp} = \sqrt{2\hbar/m \omega_{\perp}}$ is the harmonic oscillator length for the radial confinement of frequency $\omega_{\perp}$, and $\zeta(s)$ is the zeta-function. 

Since the 3D scattering volume can be tuned via a p-wave Feshbach resonance \cite{Thywissen16}, the confinement strength can tune both $\ell$ and $r$; a phenomenon called the confinement induced resonance \cite{Olshanii98, Olshanii03}. Across the confinement induced resonance $\ell$ can be tuned continuously from positive to negative infinity via the confinement induced resonance. The 1D effective range is more or less constant, but its value  can be enhanced when $a_{\perp}$ is larger than the 3D effective range, $R$. Hence the energy dependence of the 1D scattering can be enhanced by the confinement.  For example the effective range for spin-polarized $^{40}K$ is \cite{Thywissen16}: $R \approx 3.8$ nm. For a radial confinement of $\omega_{\perp} \approx 2\pi * 400$ kHz, or equivalently $a_{\perp} \approx 25$ nm, the 1D effective range is $r \approx 18$ nm, which is larger than the 3D effective range by almost a factor of $5$. In fact this effect becomes stronger for weaker radial confinement.  Although the 1D effective range is an irrelevant quantity to the energetics and dynamics in the renormalization group sense \cite{Sachdev}, the effective range may be important to the dynamics due to its enhancement from the radial confinement. The importance of the effective range then depends not only on the typical energy scale in the problem, according to the effective range expansion Eq.~(\ref{eq:effective_range}), but also this enhancement from the radial confinement. 

\begin{figure}
\includegraphics[scale=0.63]{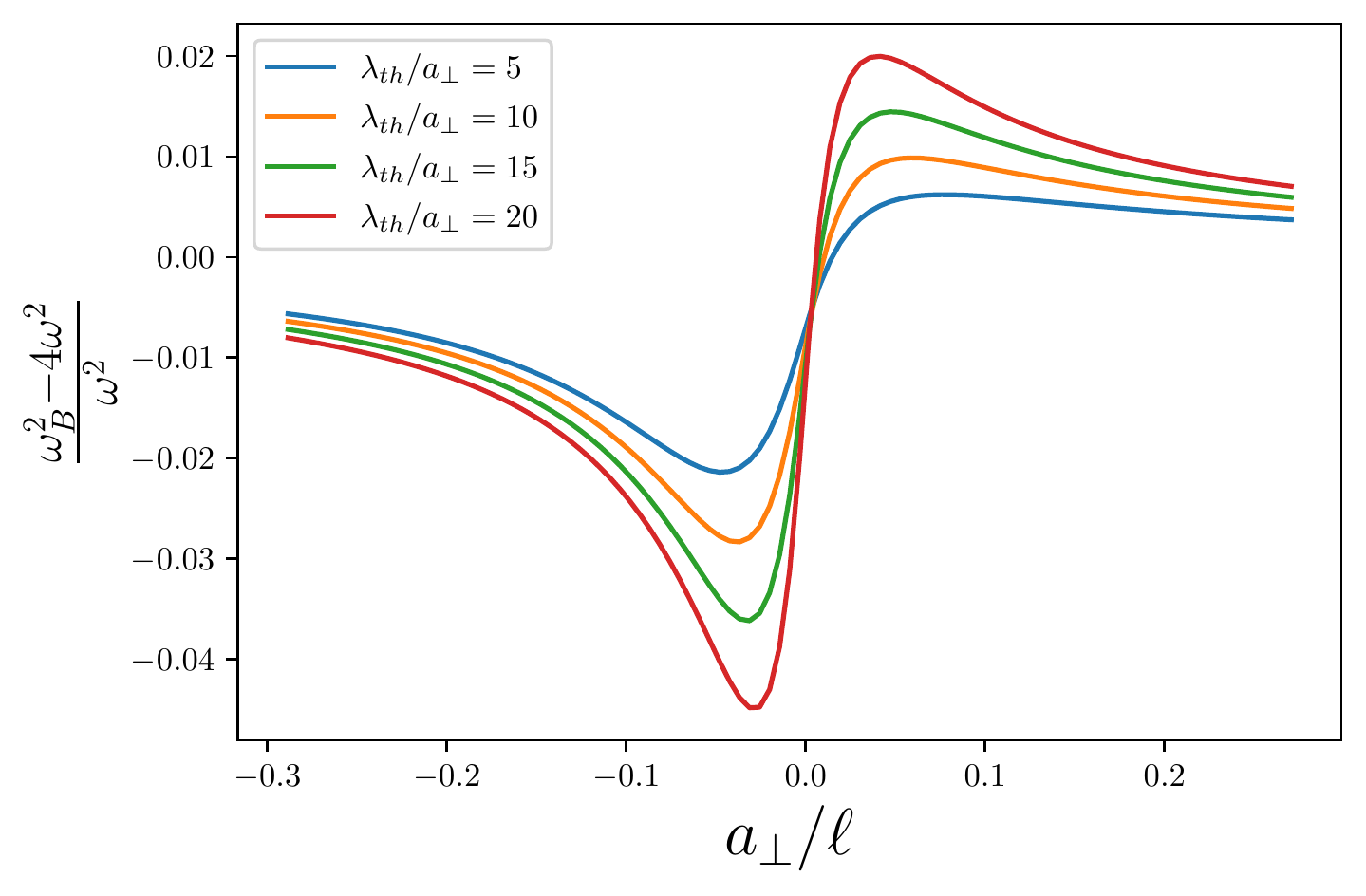}
\caption{Shift in the breathing mode frequency at various temperatures across the confinement induced resonance, Eq.~(\ref{eq:cir}), for $^{40}K$. Here $\lambda_{th}$ is the thermal de Broglie wavelength, the axial trapping frequency is $\omega = 2\pi * 200$ Hz,  the transverse harmonic length is $a_{\perp} \approx 25$ nm, $N=10$, and I have used the parametrization of the p-wave Feshbach resonance at $198.3$ G \cite{Thywissen16, Fesh_Note}.}
\label{fig:breahting_mode_T}
\end{figure}

To further examine the experimental applicability of these results, the shift in the breathing mode frequency for a 1D spin polarized Fermi gas confined in an axial harmonic potential with frequency $\omega \approx 2\pi* 200$ Hz is presented in Fig.~(\ref{fig:breahting_mode_T}). For a radial confinement with frequency $\omega_{\perp} \approx 2\pi*400$ kHz, the quasi 1D regime is defined via: $\hbar \omega_{\perp} \gg k_B T$, while the high temperature limit is given by: $T\gg N \hbar \omega$, see Appendix \ref{app:LDA}. For the parameters under consideration this leads to $40 \ \mu K > T > 50 \  nK $, where I have used $N \approx 10$. 

The results in Fig.~(\ref{fig:breahting_mode_T}) are for temperatures $5 \ \mu K > T> 0.4 \ \mu K$, which sits within the quasi 1D-regime and high-temperature limits. At resonance, the shift in the contact is small and only a few percent. However slightly away from resonance and for $\ell <0$, the shift can be about $3-5\%$ depending on the temperature. The asymmetry in Fig.~(\ref{fig:breahting_mode_T}) also implies that the effective range is important in describing the shift in the breathing mode frequency. 


\section{Conclusions}
\label{sec:conc}

The main results of this work are the evaluation of the thermodynamic contacts and the shift of the breathing mode frequencies in the high temperature limit, see Eqs.~(\ref{eq:contacts_near_res}-\ref{eq:contacts_res}, \ref{eq:breathing_mode_shift_final}-\ref{eq:shift_weak}). These results are valid to lowest non-trivial order in the virial expansion, i.e. to second order in $\left(n \lambda_{th}\right)^2$.

Although the above results extend to arbitrary interactions strengths, it is important to note that this collective mode does possess a finite lifetime when the scattering volume is finite. The finite lifetime of the breathing mode can be shown to be related to the finite bulk viscosity, $\zeta$, and scales as $\zeta^{-1}$ \cite{Maki20b}. In the high-temperature limit, the bulk viscosity will also occur at order $O\left(n \lambda_{th}\right)^2$ \cite{Maki20, Enss19, Nishida19, Hofmann20}. Moreover, since the bulk viscosity must be positive, as it is related to entropy production, the bulk viscosity ought to be proportional to $\ell^2$ and $r^2$ for a weakly interacting and resonantly interacting gas, respectively. Practically one expects the breathing mode picture breaks down around $\tilde{\ell} \approx 1$. A more in-depth study of the bulk viscosity is beyond the scope of this work and will be presented in the future.

One interesting part of this study is that the effective range for a 1D spin-polarized Fermi gas can be enhanced by the presence of radial confinement, see Eq.~(\ref{eq:cir}). This is due to the competition between the validity of the low-energy three-dimensional scattering, set by $R$, and the validity of the quasi-1D regime, set by $a_{\perp}$. This increase of the effective range can lead to deviations from the zero-range limit at smaller energy scales. This is evident in Fig.~(\ref{fig:breahting_mode_T}). There the contribution to the shift in the breathing mode frequency due to the effective range is important for $\ell <0$. A similar phenomenon has also been emphasized for the case of quasi two-dimensional Fermi superfluids in Ref.~\cite{Liu19b}. 

This situation ought to be contrasted to the spin 1/2 s-wave Fermi gas in 3D. This system has the same renormalization group structure as the spin-polarized 1D Fermi gas with p-wave interactions \cite{Sachdev}. In both cases, the effective range is an irrelevant quantity, and leads to perturbative effects; a conclusion consistent with effective field theory \cite{Kolck99, Kolck02}. As the effective range is irrelevant, the 3D s-wave Fermi gas has a well defined zero-range limit, where the physics are universally described by the s-wave scattering length. In this case, the effective range can often be ignored as it is generally quite small. For the spin-polarized 1D Fermi gas, although a well-defined zero-range limit exists, the effective range can be important due to its enhancement from the transverse confinement, as evidenced by Figs.~(\ref{fig:breathing_mode_ell}-\ref{fig:breahting_mode_T}).

I would like to thank Shizhong Zhang and Joseph H. Thywissen for useful discussions. This work was supported through the Research Grants Council of the Hong Kong Special Administrative Region, China (General research fund, HKU 17304719, 17304820 and collaborative research fund C6026-16W and C6005-17G), and the Croucher Foundation under the Croucher Innovation Award.

\appendix

\numberwithin{equation}{section}
\renewcommand\theequation{\Alph{section}.\arabic{equation}}

\section{Derivation of the Two-Body T-Matrix}
\label{app:T_matrix}

Here I evaluate the two-body T-matrix in the presence of the vacuum. The two-body T-matrix for  identical fermions depends on the dressed propagator for the closed channel molecular state. The equation for the dressed molecular propagator is:

\begin{align}
D^{-1}(Q,Q_0) &= D^{-1}_0(Q,Q_0) - \Sigma(Q,Q_0) \nonumber \\
D^{-1}_0(Q,Q_0) &= Q_0 - \frac{1}{4}Q^2 -\nu_0 + i\delta
\label{app:A0:D}
\end{align}

\noindent Pictorially, this equation is shown in Fig.~(\ref{fig:t-matrix}). In Eq.~(\ref{app:A0:D}) $\Sigma(Q,Q_0)$ is the fermion-fermion bubble which is given by:

\begin{align}
\Sigma(Q,Q_0) &= -\frac{g^2}{2} \int_{-\infty}^{\infty} \frac{dk}{2\pi}  \int_{-\infty}^{\infty} \frac{dk_0}{2\pi i} \  k^2 \nonumber \\
& \left[ \frac{1}{\frac{Q_0}{2}+k_0 - \frac{1}{2}\left(\frac{Q}{2}+k\right)^2+ i \delta}\right. \nonumber \\
&\left. \frac{1}{\frac{Q_0}{2}-k_0 - \frac{1}{2}\left(\frac{Q}{2}-k\right)^2+i \delta} \right]\nonumber \\
&= \frac{g^2}{2} \int_{-\infty}^{\infty} \frac{dk}{2\pi}\frac{1}{Q_0 - \frac{1}{4}Q^2 - k^2 + i\delta}
\label{app:A0:sigma}
\end{align}

\noindent In defining Eqs.~(\ref{app:A0:D}-\ref{app:A0:sigma}), I have ignored the many-body background by setting the chemical potential to zero.

One can see from Eq.~(\ref{app:A0:sigma}) that the integrand is divergent at large momenta. To this end, add a ultraviolet cut-off scale, $\Lambda$ to regularize the integrals. A direct evaluation of Eq.~(\ref{app:A0:sigma}) then gives:

\begin{align}
\Sigma(Q,Q_0) &= -\frac{g^2}{2\pi}\Lambda + \frac{g^2}{4} \sqrt{-Q_0 + \frac{Q^2}{4} - i \delta} \nonumber \\
&= -\frac{g^2}{2L} \sum_k + \frac{g^2}{4}\sqrt{-Q_0 + \frac{Q^2}{4} - i \delta}
\end{align}

\noindent The resulting dressed molecular propagator is then given by:

\begin{align}
D^{-1}(Q_0,Q)  &= -\nu_0 + \frac{g^2}{2L} \sum_k + Q_0 - \frac{1}{4}Q^2 \nonumber \\
&- \frac{g^2}{4}\sqrt{-Q_0 + \frac{Q^2}{4}- i \delta}
\label{app:A0:Dfinal}
\end{align}

\noindent where I have written $\Lambda/\pi = L^{-1}\sum_k $. Eq.~(\ref{app:A0:Dfinal}) is equivalent to Eq.~(\ref{eq:vacuum_T-matrix}).

\section{Derivation of the Contacts}
\label{app:contacts}

In this appendix I show how to obtain the microscopic definitions of the contact operators, and how they relate to the pressure. First note that the pressure can be written as:

\begin{equation}
P = \frac{1}{\beta^{3/2}} G\left(\beta \mu, \beta^{1/2} \frac{1}{\ell}, \beta^{-1/2} r\right)
\label{app:A:pressure_rel_1}
\end{equation}

\noindent where $\beta = 1/T$ is the temperature, $\mu$ the chemical potential, $\ell$ the scattering volume, $r$ the effective range, and finally $G$ is a dimensionless function. In writing Eq.~(\ref{app:A:pressure_rel_1}) I have used the fact both $\ell$ and $r$ have dimensions of length. More formally, the pressure is defined in terms of the partition function, $Z$, via:

\begin{equation}
P = \frac{1}{\beta L} \ln\left(Z\right) = \frac{1}{\beta L} \ln\left(Tr\left[e^{-\beta (H-\mu N)}\right]\right)
\label{app:A:pressure_def}
\end{equation}

From Eqs.~(\ref{app:A:pressure_rel_1}-\ref{app:A:pressure_def}), and by considering the derivative of $P$ with respect to $\beta$, one can show that the pressure must satisfy:

\begin{equation}
P L = 2 \langle H \rangle + \frac{1}{\ell}\langle C_{\ell}\rangle - r \langle C_r \rangle
\label{app:A:pressure_rel_2}
\end{equation}

\noindent where $\langle \cdot  \rangle$ denotes a thermal average. In Eq.~(\ref{app:A:pressure_rel_2}), the thermal average of the scattering volume contact, $C_{\ell}$, and the effective range contact, $C_r$, are defined via:

\begin{align}
\langle C_{\ell} \rangle &= \left\langle - \frac{\partial H}{\partial \ell^{-1}} \right\rangle \nonumber \\
\langle C_{r} \rangle &= \left\langle - \frac{\partial H}{\partial r} \right\rangle
\label{app:A:contacts}
\end{align}

To progress further it is important to note that the scattering quantities $\ell$ and $r$ are functions of the bare quantities $\nu_0$ and $g$ in the Hamiltonian, see Eq.~(\ref{eq:scattering_parameters}). Evaluating Eq.~(\ref{app:A:contacts}), one can find the following microscopic definitions for the contact operators:

\begin{align}
 C_{\ell}  &= \int_{-\infty}^{\infty} dx \frac{g^2}{4} \phi^{\dagger}(x) \phi(x)\nonumber \\
C_{r}&= \int_{-\infty}^{\infty} dx \frac{g^2}{4} \phi^{\dagger}(x,t) \left(i \partial_t + \frac{\partial_x^2}{4}+ 2 \mu\right)\phi(x,t) 
\label{app:A:contacts_final}
\end{align}

\noindent In order to obtain Eq.~(\ref{app:A:contacts_final}), I have used the Heisenberg equation of motion for the molecular field: $\partial_t \phi(x,t) = i \left[H, \phi(x,t)\right]$, to integrate out the fermionic degrees of freedom in the definition of $C_r$.

\section{Imaginary Part of the T-Matrix}
\label{app:boundstate}

In this appendix I calculate the imaginary part of the T-matrix to leading order in the virial expansion. There are two contributions, the first is due to the positive frequency scattering states, and the second is due to the bound state.

To begin, the molecular propagator to leading order in the virial expansion is given by:

\begin{align}
&\left(\frac{g^2D(Q,z-i \delta)}{4}\right)^{-1} = \frac{1}{\ell} + \left(z-\frac{Q^2}{4}+ 2 \mu\right) r \nonumber \\
&+ \int_{-\infty}^{\infty} \frac{dk}{\pi}\ k^2  \left[\frac{1}{k^2 - z + \frac{Q^2}{4}- 2\mu +i \delta} - \frac{1}{k^2}\right] \nonumber \\
&= \frac{1}{\ell} + \left(z-\frac{Q^2}{4}+2\mu\right) - \sqrt{-z +\frac{Q^2}{4} -2\mu + i \delta}
\label{app:B:T_matrix}
\end{align}

\noindent which is simply the molecular propagator in the presence of the vacuum. From Eq.~(\ref{app:B:T_matrix}) one can absorb the $Q$ and $\mu$ dependence by shifting variables to $z' = z - Q^2/4+2\mu$. Hence I will ignore the dependence of the $T$-matrix on $Q$ and $\mu$. For positive frequencies, $z>0$, one can show that the imaginary part of the $T$-matrix is:

\begin{equation}
\Im \left[ \frac{g^2D(0,z-i \delta)}{4}\right] = \frac{\sqrt{z}}{\left(\frac{1}{\ell} + z r\right)^2 + z}
\end{equation}

At negative frequencies, $z<0$, the imaginary part of the $T$-matrix comes from the two-body bound state. The two-body bound state for this system is given by the pole of the $T$-matrix, or equivalently the pole of Eq.~(\ref{app:B:T_matrix}). This occurs when:

\begin{equation}
\left(\frac{g^2D(0,-z_B)}{4}\right)^{-1} = 0 =  \frac{1}{\ell} - z_B r - \sqrt{z_B}
\label{app:B:boundstate}
\end{equation}

As one can see from Eq.~(\ref{app:B:boundstate}), the low-energy bound state exists for $\ell >0$. For small effective range, $r^2 z_B \ll 1$, one obtains: 

\begin{equation}
z_B = \frac{1}{\ell^2} \left(1 - 2\frac{r}{\ell}\right)
\end{equation}

\noindent which is perturbative in the effective range, $r$. This is evidence for the statement that the effective range is an irrelevant quantity in understanding the energetics and dynamics. 

In principle Eq.~(\ref{app:B:boundstate}) also has a pole that is approximately given by: $z_B \approx 1/r^2$. This term must be discarded as it is beyond the low-energy approximation which requires: $z_B \ll r^{-2}$.

To evaluate the imaginary part of the $T$-matrix for negative frequencies, expand the denominator of the $T$-matrix around $-z_B$:

\begin{equation}
\left[\frac{g^2D(0,z- i \delta)}{4}\right]^{-1} \approx \left(\frac{1}{2\sqrt{z_B}} + r\right)(z+z_B) - i \delta
\end{equation}

\noindent The imaginary part of the $T$-matrix at negative frequencies is then:

\begin{equation}
\Im \left[\frac{g^2 D(Q, z-i \delta)}{4}\right] = \left(\frac{1}{2 \sqrt{z_B}} + r\right)^{-1} \pi \delta(z + z_B) \theta(\ell)
\end{equation}

In terms of the contacts, the bound state contributions to Eq.~(\ref{eq:contacts_2}) are equal to:

\begin{align}
\left. \frac{\langle C_{\ell}\rangle}{L}\right|_{\text mol.} &= \frac{e^{2\beta \mu}}{\sqrt{\pi \beta}}  e^{\beta z_B} \frac{2\sqrt{z_B}}{1+ 2 \sqrt{z_B}r} \theta(\ell) \nonumber \\
\left. \frac{\langle C_{r}\rangle}{L}\right|_{\text mol.} &= -\frac{e^{2\beta \mu}}{\sqrt{\pi \beta}}\ e^{\beta z_B}z_B \frac{2\sqrt{z_B}}{1+ 2 \sqrt{z_B}r} \theta(\ell)
\label{app:B:eq:contacts}
\end{align}

 From Eq.~(\ref{app:B:eq:contacts}), one can see that the molecular contribution vanishes at resonance, while it appears to become quite large in the weakly interacting limit. Near resonance, and for $\ell>0$ the bound state energy is small but finite. In this limit the contribution due to the closed channel molecules is important. Near resonance there can be a large population of molecules since the wavefunctions of the closed channel molecules have substantial overlap with the scattering states. In the weakly interacting limit, however, the bound state is exceptionally deep. For a thermal gas in the upper branch \cite{Feshbach_Res}, the relaxation of scattering atoms into the closed channel molecules is exceptionally slow.  For this reason it is safe to neglect the molecular contribution to the contacts.


\section{Local Density Approximation at High-Temperatures}
\label{app:LDA}

In this appendix I discuss how to evaluate the contacts for a trapped 1D Fermi gas in the local density approximation at high temperatures. At high-temperatures, the Fermi-Dirac distribution can be replaced by a Boltzmann distribution:

\begin{equation}
f(k,x) = \exp\left[-\beta \left(\frac{k^2}{2} + \frac{1}{2} \omega^2 x^2 -\mu \right)\right]
\end{equation}

\noindent where I have set $\hbar$ and $m$ to be unity for the time being and $\mu$ is the chemical potential at the center of the trap. The density is then given by:

\begin{equation}
n(x) = \int_{-\infty}^{\infty} \frac{dk}{2\pi} f(k,x) = 
\frac{1}{\sqrt{2\pi \beta}} \exp\left[-\beta\left(\frac{1}{2}  \omega^2 x^2 -\mu\right)\right]
\end{equation}

\noindent normalizing with respect to the total number of particles, $N$ gives the following condition for the chemical potential:

\begin{equation}
N = \frac{e^{\beta \mu}}{\beta \omega}
\end{equation}

\noindent or equivalently the density is given by:

\begin{equation}
n(r) = \frac{\beta \hbar \omega N}{\lambda_{th}}e^{-\frac{1}{2}\beta m \omega^2 x^2}
\label{eq:lda_density}
\end{equation}

\noindent where I have restored the factors of $\hbar$ and $m$. Note here that the virial expansion is equivalent to the statement $N \beta \hbar \omega \ll 1$.

From Eq.~(\ref{eq:lda_density}) the initial moment of inertia can then be calculated:

\begin{align}
\langle x^2 \rangle(0) &= \int_{-\infty}^{\infty} dx x^2 n(x) =  \frac{N}{m \beta \omega^2}
\end{align}


\begin{thebibliography}{References}
\bibitem{Pressure_Note} The average of the trace of the stress-energy tensor is related to the pressure, $P$, via:  $\langle \Pi \rangle = P L$ for one dimensional systems, where $L$ is the length of the system. Eq.~(\ref{eq:operators}) then follows from the results of Appendix~\ref{app:contacts}.
\bibitem{Fesh_Note} In experiments the 3D p-wave Feshbach resonance is anisotropic. However, for simplicity I treat it as isotropic.
\end{thebibliography}
\end{document}